%% file: main.tex
\documentclass[sigconf,authorversion,nonacm]{acmart}

\AtBeginDocument{%
  \providecommand\BibTeX{{%
    \normalfont B\kern-0.5em{\scshape i\kern-0.25em b}\kern-0.8em\TeX}}}

\setcopyright{acmcopyright}
\copyrightyear{2022}
\acmYear{2022}

\acmConference[ICSE 2022]{The 44th International Conference on Software Engineering}{May 21--29, 2022}{Pittsburgh, PA, USA}
\acmBooktitle{44th International Conference on Software Engineering (ICSE '22), May 21--29, 2022, Pittsburgh, USA}
\acmPrice{15.00}
\acmISBN{XXXX}

\usepackage{algorithmic}
\usepackage{graphicx}
\usepackage{textcomp}
\usepackage{xcolor, colortbl}
\usepackage{booktabs}
\usepackage{multirow}
\usepackage{xspace}
\usepackage{caption}
\usepackage{subcaption}
\usepackage{footnote}
\usepackage{url}
\usepackage{tcolorbox}
\usepackage{booktabs}
\usepackage{enumitem}
\usepackage{fixltx2e}
\usepackage{multicol}
\usepackage{balance}
\usepackage{listings}
\usepackage[normalem]{ulem}

\definecolor{Gray}{gray}{0.9}
\definecolor{codegreen}{rgb}{0,0.6,0}
\definecolor{codegray}{rgb}{0.73,0.38,0.06}
\definecolor{codepurple}{rgb}{0.70,0.27,0}
\definecolor{codemagenta}{rgb}{0.74,0.09,0.42}
\definecolor{codeoutput}{rgb}{0.5,0,0}
\definecolor{backcolour}{rgb}{0.96,0.96,0.96}

\def\BibTeX{{\rm B\kern-.05em{\sc i\kern-.025em b}\kern-.08em
    T\kern-.1667em\lower.7ex\hbox{E}\kern-.125emX}}

\newboolean{showcomments}

\setboolean{showcomments}{true}

\ifthenelse{\boolean{showcomments}}
  {\newcommand{\nb}[2]{
    \fbox{\bfseries\sffamily\scriptsize#1}
    {\sf\small$\blacktriangleright$\textit{#2}$\blacktriangleleft$}
   }
  }
  {\newcommand{\nb}[2]{}
  }

\definecolor{lightergray}{rgb}{0.9,0.9,0.9}
\newtcolorbox{resultbox}{colback=lightergray, arc=0.5mm, top=2mm, bottom=2mm, left=2mm, right=2mm}

\newcommand{\ie}{\emph{i.e.,}\xspace}
\newcommand{\eg}{\emph{e.g.,}\xspace}
\newcommand{\etc}{etc.\xspace}
\newcommand{\etal}{\emph{et~al.}\xspace}
\newcommand{\secref}[1]{Section~\ref{#1}\xspace}
\newcommand{\figref}[1]{Fig.~\ref{#1}\xspace}

\newcommand{\tabref}[1]{Table~\ref{#1}\xspace}

\newcommand{\approach}{{\sc \emph{RELINE}}\xspace}
\newcommand{\random}{\emph{random agent}\xspace}
\newcommand{\baseline}{\emph{rl-baseline}\xspace}

\newcommand{\gameOne}{CartPole\xspace}
\newcommand{\gameTwo}{MsPacman\xspace}

\title[Using Reinforcement Learning for Load Testing of Video Games]{Using Reinforcement Learning for\\Load Testing of Video Games}

\author{Rosalia Tufano}
\affiliation{%
  \institution{SEART @ Software Institute\\Universit\`a della Svizzera italiana}
  \country{Switzerland}
}
\author{Simone Scalabrino}
\affiliation{%
  \institution{STAKE Lab\\University of Molise}
  \country{Italy}
}
\author{Luca Pascarella}
\affiliation{%
  \institution{SEART @ Software Institute\\Universit\`a della Svizzera italiana}
  \country{Switzerland}
}
\author{Emad Aghajani}
\affiliation{%
  \institution{SEART @ Software Institute\\Universit\`a della Svizzera italiana}
  \country{Switzerland}
}
\author{Rocco Oliveto}
\affiliation{%
  \institution{STAKE Lab\\University of Molise}
  \country{Italy}
}
\author{Gabriele Bavota}
\affiliation{%
  \institution{SEART @ Software Institute\\Universit\`a della Svizzera italiana}
  \country{Switzerland}
}

\copyrightyear{2022} 
\acmYear{2022} 
\setcopyright{acmcopyright}\acmConference[ICSE '22]{44th International Conference on Software Engineering}{May 21--29, 2022}{Pittsburgh, PA, USA}
\acmBooktitle{44th International Conference on Software Engineering (ICSE '22), May 21--29, 2022, Pittsburgh, PA, USA}
\acmPrice{15.00}
\acmDOI{10.1145/3510003.3510625}
\acmISBN{978-1-4503-9221-1/22/05}

\begin{document}

\begin{abstract}
Different from what happens for most types of software systems, testing video games has largely remained a manual activity performed by human testers. This is mostly due to the continuous and intelligent user interaction video games require. Recently, reinforcement learning (RL) has been exploited to partially automate functional testing. RL enables training \emph{smart} agents that can even achieve super-human performance in playing games, thus being suitable to explore them looking for bugs. We investigate the possibility of using RL for load testing video games. Indeed, the goal of game testing is not only to identify functional bugs, but also to examine the game's performance, such as its ability to avoid lags and keep a minimum number of frames per second (FPS) when high-demanding 3D scenes are shown on screen. We define a methodology employing RL to train an agent able to play the game as a human while also trying to identify areas of the game resulting in a drop of FPS. We demonstrate the feasibility of our approach on three games. Two of them are used as proof-of-concept, by injecting artificial performance bugs. The third one is an open-source 3D game that we load test using the trained agent showing its potential to identify areas of the game resulting in lower FPS.
\end{abstract}

\maketitle


\input{introduction}

\input{approach}
\input{study1}
\input{study2}

\input{threats}
\input{related}
\input{conclusion}

\section*{Acknowledgment}
This project has received funding from the European Research Council (ERC) under the European Union's Horizon 2020 research and innovation programme (grant agreement No. 851720). Any opinions, findings, and conclusions expressed herein are the authors' and do not necessarily reflect those of the sponsors. 


\bibliographystyle{ACM-Reference-Format}
\bibliography{main}

\end{document}

%% file: introduction.tex


\section{Introduction}
\label{sec:intro}
The video game market is expected to exceed \$200 billion in value in 2023 \cite{market}. In such a competitive market, releasing high-quality games and, consequently, ensuring a great user experience, is fundamental. However, the unique characteristics of video games (from hereon, games) make their quality assurance process extremely challenging. Indeed, besides inheriting the complexity of software systems, games development and maintenance require a diverse set of skills covered by many stakeholders, including graphic designers, story writers, developers, AI (Artificial Intelligence) experts, \etc \eject

Also, games can hardly benefit from testing automation techniques \cite{Pascarella:msr2018}, since even just exploring the total space available in a given game level requires an \emph{intelligent} interaction with the game itself. For example, in a racing game, identifying a bug that manifests when the finish line is crossed requires a player able to successfully drive the car for the whole track (\ie requires the ability to drive the car). Thus, random exploration is not a viable option here.  

Therefore, it comes without surprise that game testing is largely a manual process. Zheng \etal \cite{Zheng:ase2019} report that 30 human testers were employed for testing one of the games used in their study. Also, the challenges in testing games have been stressed by Lin \etal \cite{Lin2016StudyingTU}, who showed that 80\% of the 50 popular games they studied have been subject to urgent updates.

To support developers with game testing, researchers have proposed several techniques. These include approaches to test the stability of game servers (\eg by generating high packet loads) \cite{Jung:2005,Lim:2006,Cho:2010}, model-based testing \cite{Iftikhar:models2015} using domain modeling for representing the game and UML state machines for behavioral modeling, as well as techniques specifically designed for testing board games \cite{Smith:2009,Mesentier:2017}. When looking at recent techniques aimed at proposing more general testing frameworks, those exploiting Reinforcement Learning (RL) are on the rise. This is due to the remarkable results achieved by RL-based techniques in playing games with super-human performance reported in the literature \cite{baker2019emergent,berner2019dota,hessel2018rainbow,mnih2013playing,mnih2015human,Vinyals2017StarCraftIA}. 

RL is a machine learning technique aimed to train \emph{smart} agents able to interact with a given environment (\eg a game) and to take decisions to achieve a goal (\eg win the game). RL is based on the simple idea of trial and error: The agent performs actions in the environment (of which it only has a partial representation) and receives a \emph{reward} that allows it to assess its past actions/behavior with respect to the desired goal.

Recently, researchers started using RL not only to play games but also to test them and, in general, to improve their quality. The common idea behind these approaches is to reduce the human effort in playtesting (\ie the process of testing a new game to look for bugs before releasing it to the market) using intelligent agents. RL-based agents have been used to help game designers, for example, in balancing crucial parameters of the game (\eg power-up item effects, characters abilities) \cite{Zhao2019,pfau2020dungeons,Zook2014AutomaticPF} and in testing the game difficulty \cite{gudmundsson2018human, Stahlke2020ArtificialPI}. Also, RL-based agents have been used to look for bugs in games \cite{Pfau:2017,Bergdahl2020,Zheng:ase2019,Ariyurek:tg2021}. 

While agents are usually trained to play a game with the goal of winning, the aforementioned works trained the agent to not only advance in the game but also to explore it to search for bugs. For example, Ariyurek \etal \cite{Ariyurek:tg2021} combine RL and Monte Carlo Tree Search (MCTS) to find issues in the behavior of a game, given its design constraints and game scenario graph (provided by the game developer). The \emph{ICARUS} framework \cite{Pfau:2017} is able to identify crashes and blockers bugs (\eg the game get stuck for a certain amount of time) while the agent is playing. Similarly, the approach by Zheng \etal \cite{Zheng:ase2019}, also exploiting RL, can identify bugs spotted by the agent during training (\eg crashes). 

While these approaches pioneered the use of RL for game testing, they are mostly aimed at testing functional (\eg finding crashes) or design-related (\eg level design) aspects. However, these are not the only types of bug developers look for in playtesting. 

In a recent survey, Politowski \etal \cite{politowski2021survey} reported that for two out of the five games they considered (\ie \textit{League of Legends} by Riot and \textit{Sea of Thieves} by Rare) developers partially automated game performance checks (\eg frame-rate). Similarly, Naughty Dog used specialized profiling tools\footnote{\url{https://youtu.be/yH5MgEbBOps?t=3494}} for finding which parts of a given scene caused a drop in the number of frames per second (FPS) in \textit{The Last of Us}. Truelove \etal \cite{truelove2021we} report that game developers agree that \textit{Implementation response} problems (among which, performance-related ones) may severely impact the game experience. Also, Li \etal \cite{li2021data} observed that players frequently complain about performance issues in game reviews. 

\textbf{Significance of research contribution.} Despite such a strong evidence about the importance of detecting performance issues in video games, to the best of our knowledge no previous work introduced automated approaches for load testing video games. 
We present \approach (\textbf{R}einforcement l\textbf{E}arning for \textbf{L}oad test\textbf{IN}g gam\textbf{E}s), an approach exploiting RL to train agents able to play a given game while trying to load test it with the goal of minimizing its FPS. The agent is trained using a \emph{reward} function enclosing two objectives: The first aims at teaching the agent how to advance in (and possibly win) the game. The second rewards the agent when it manages to identify areas of the game exhibiting low FPS. The output of \approach is a report showing to developers the identified areas in the game being negative outliers in terms of FPS, accompanied by videos of the gameplays exhibiting the issue. Such ``reports'' can simplify the identification and reproduction of performance issues, that are often reported in open-source 3D games (see \eg \cite{dwarfcorp3,3dcity,geostrike,dwarfcorp4}) and that, in some cases, are challenging to reproduce (see \eg \cite{dwarfcorp1,dwarfcorp2}), even requiring special instructions for their reporting \cite{dwarfcorp5}. \smallskip

We experiment \approach with three games. The first two are simple 2D games that we use as a proof-of-concept. In particular, we injected in the games artificial ``performance bugs'' \cite{performance-mutation} to check whether the agent is able to spot them. We show that the agent trained using \approach can identify the injected bugs more often than (i) a random agent, and (ii) a RL-based agent only trained to play the game. Then, we use \approach to load test an open-source 3D game \cite{kart}, showing its ability to identify areas of the game being negative outliers in terms of FPS.

Code and data from our study are publicly available \cite{replication}.

%% file: approach.tex

\suppressfloats

\begin{figure}[t]
	\centering
	\includegraphics[width=\linewidth]{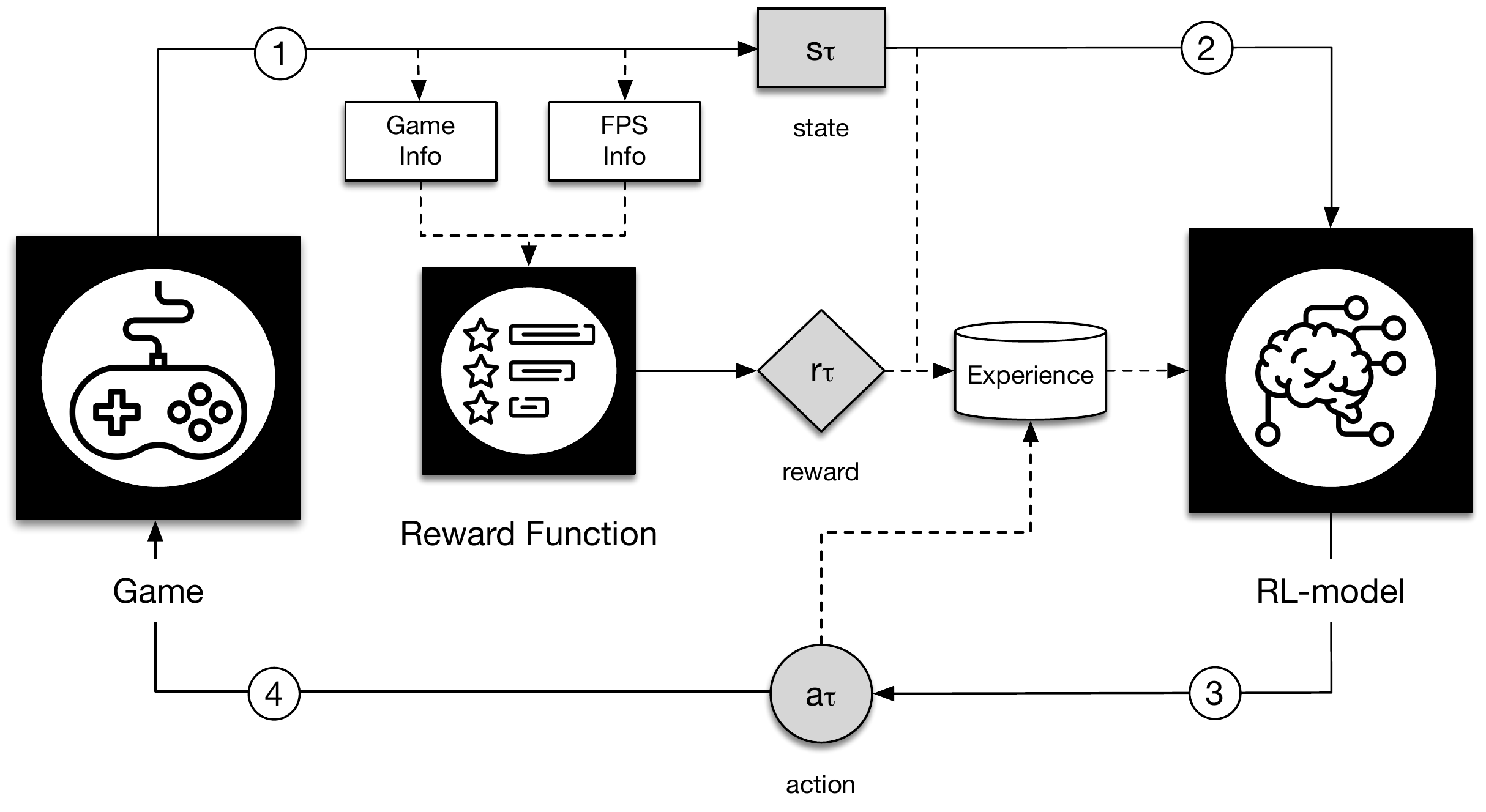}
	\caption{\emph{RELINE} overview}
	\label{fig:approach}
	\vspace{-0.3cm}
\end{figure}

\section{RL to Load Test Video Games}
\label{sec:approach}

In this section we explain, from an abstract perspective, the idea behind \approach. We describe in the study designs how we instantiated \approach to the different games we experiment with (\eg details about the adopted RL models). 

\approach requires three main components: the \emph{game} to load test, a \emph{RL model}, representing the agent that must learn how to play the game while load testing it, and a \emph{reward function}, used to reward the agent so that it can evaluate the worth of its actions for reaching the desired goal (\ie playing while load testing).
The \emph{RL model} is trained through the 4-step loop depicted in \figref{fig:approach} (see the circled numbers). The continuous lines represent steps performed at each iteration of the loop, while the dashed ones are only performed after a first iteration has been run (\ie after the agent performed at least one action in the game). When the first episode (\ie a run of the game) of the training starts (step 1), at each time step $\tau$ the game provides its state $s_\tau$. Such a state can be, for example, a set of frames or a numerical vector representing what is happening in the game (\eg the agent's position). The \emph{RL model} takes as input $s_\tau$ (step 2) and provides as output the action $a_\tau$ to perform in the game (step 3). When the agent has no experience in playing the game at the start of the training, the weights of the neural network in the RL model are randomly initialized, producing random actions.
The action $a_\tau$ is executed in the game (step 4), which, in turn, generates the subsequent state $s_{\tau + 1}$. 

After the first iteration (\ie after having received at least one $a_\tau$), the game also produces, at each iteration, the data needed to compute the reward function. In \approach we collect (i) the information needed to assess how well the agent is playing the game (\eg time since the episode started and the episode score), and (ii) the FPS at time $\tau$. It is required that the game developer instruments the game and provide APIs through which \approach can acquire such pieces of information. We assume that this requires a minor effort.

The \emph{reward function} aims at training an agent that is able to (i) play the game, thanks to the information indicating how well the agent is playing, and (ii) identify low-FPS areas, thanks to the information about the FPS. 
The output of the \emph{reward function} is a number representing the reward obtained by the agent at time $\tau$. In \approach, the reward function for a given action is composed of two sub-functions: A \textit{game reward function}, depending on how good the action is in the game ($\mathit{rg}_\tau$), and a \textit{performance reward function}, depending on how the action impacts the performance ($\mathit{rp}_\tau$). 

We combine such functions in $r_\tau = \mathit{rg}_\tau + \mathit{rp}_\tau$.
The game reward function clearly depends on the game under test: A function designed for a racing game likely makes no sense for a role-playing game. In general, defining the reward function for learning to play should be performed by considering (i) what the goal of the game is (\eg drive on a track), and (ii) which information the game provides about the ``successful behavior of the player'' (\eg is there a score?). Even if less intuitive, the performance reward function is game-dependent as well: Assuming a tiny FPS drop (\eg -1\%), the reward can be small for a role-playing game, in which it likely does not affect the whole experience, while it should be high for an action game, in which it could even cause the (unfair) player's defeat. Unlike the game reward function, we expect however minor changes to be required to adapt the performance reward function to a different video game (\ie tuning of the thresholds to use).

The state $s_\tau$, the action $a_\tau$, and the reward $r_\tau$ are then stored in an experience buffer. When enough experience has been accumulated, it is used to update the network weights. How experience is stored and used to update the network depends on the used RL model. 

The episode ends when a final state is reached. Again, the definition of the final state depends on the game, and it could be based on a timeout (\eg each episode lasts at most 90 seconds) or on a specific condition that must be met (\eg the agent crosses the finish line). Once the episode ends, the game is reinitialized and the loop restarts. The training is performed for a number of episodes sufficient to observe a convergence in the total reward achieved by an agent during an episode (\eg if the trained agent obtains a reward of 100 for ten consecutive episodes the training is stopped).

%% file: study1.tex

\section{Preliminary Study: Injecting Artificial Performance Issues}
\label{sec:study1}

This preliminary study aims at assessing the ability of \approach in identifying artificial ``performance bugs'' \cite{performance-mutation} we simulate in two 2D games. It is important to highlight that the \emph{goal} of this study is only to demonstrate the applicability of \approach for load testing games as a proof-of-concept. A case study on a 3D open-source game is presented in \secref{sec:study2}.

\begin{figure}[t]
	\centering
	\includegraphics[width=\linewidth]{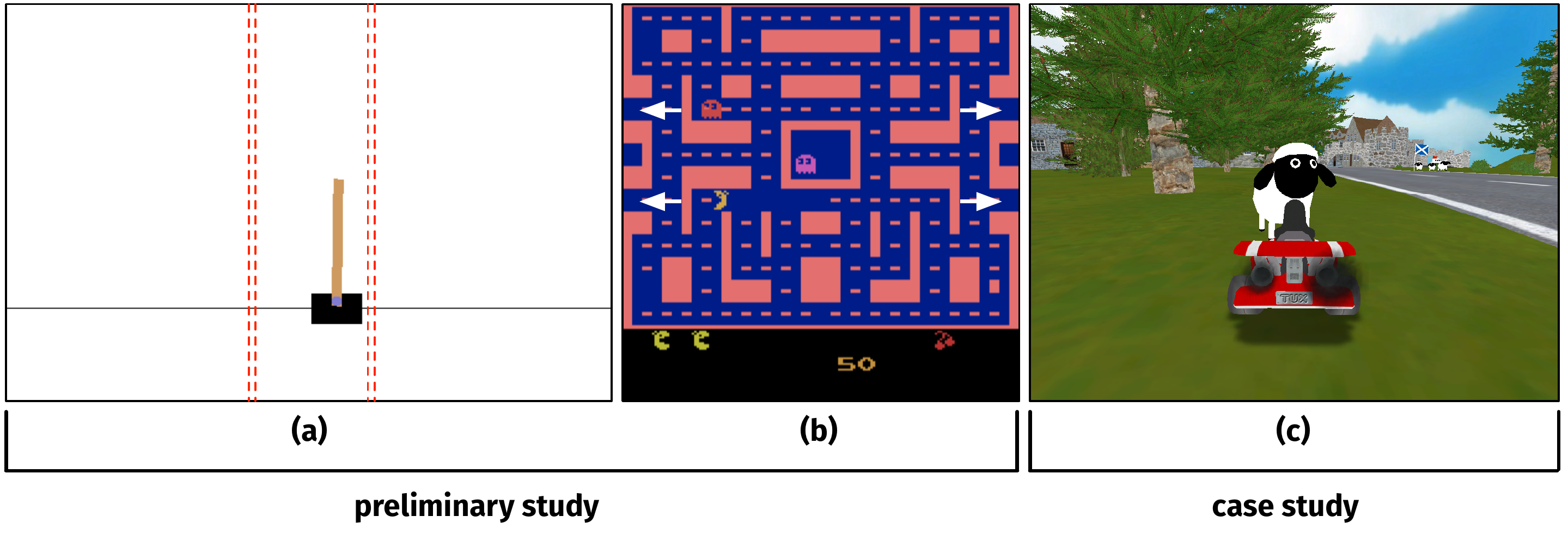}
	\caption{Screenshots of games used in the preliminary study---\secref{sec:study1} (a) \gameOne and (b) \gameTwo, and in the case study---\secref{sec:study2} (c) SuperTuxKart.}
	\label{fig:games}
\end{figure}

\subsection{Study Design}
We select two 2D games, \gameOne~\cite{CartPoleLink} and \gameTwo~\cite{PacmanLink}. The former --- \figref{fig:games}-(a) --- is a dynamic system in which an unbalanced pole is attached to a moving cart, and the player must move the cart to balance the pole and keep it in a vertical position. 

The player loses if the pole is more than 12 degrees from vertical or the cart moves too far from the center. The latter --- \figref{fig:games}-(b) --- is the classic Pac-Man game in which the goal is to eat all dots without touching the ghosts. Both games employ simple 2D graphics which bound the player's possible moves in only one (\eg left and right, for \gameOne) or two (\eg left, right, up, and down, for \gameTwo) dimensions. This is one of the reasons we selected these games for assessing whether a RL-based agent that learned how to play them can also be trained to look for artificial ``performance bugs'' we injected. Also, both games are integrated in the popular \textsc{Gym} Python toolkit~\cite{GymToolkit} developed by OpenAI~\cite{GymOfficialRef}.

\textsc{Gym} can be used for developing and comparing RL-based agents in playing games. It acts as a middle layer between the \textit{environment} (the game) and the \textit{agent} (a virtual player). In particular, \textsc{Gym} collects and executes \textit{actions} (\eg go left, go right) generated by the agent and returns to it the new state of the \textit{environment} (\ie screenshots) with additional information such as the score in the episode. \textsc{Gym} comes with a set of integrated arcade games including the two we used in this preliminary study.

\subsubsection{Bug Injection} We injected two artificial ``performance bugs'' in \gameOne and four in \gameTwo. The idea behind them is simple: When the agent visits specific areas for the first time during a game, the bugs reveal themselves (simulation of heavy resource loading). A natural way of achieving this goal would have been to introduce the bugs in the source code of the game and to implement the logic to spot FPS drops in the agent accordingly. This, however, would have slowed down the training of the agent. Therefore, we chose to use a more practically sound approach, inspired by the simulation of Heavy-Weight Operation (HWO) operator for performance mutation testing \cite{performance-mutation}: We directly assume that the agents observe the bugs when they visit the designated areas and act accordingly.

In \gameOne, the agent can only move on the $x$ axis (\ie left or right). When the game starts, the agent is in position $x=0$ (\ie center of the axis) and it can change its position towards positive (by moving right) or negative (left) $x$ values. The two bugs we injected manifest when $x \in [-0.50, -0.45]$ and $x \in [0.45, 0.50]$ --- dashed lines in \figref{fig:games}-(a). We use intervals rather than specific values (\eg -0.45) because the position of the agent is a float: if it moves to position -0.450001, we want to reward it during the training for having found the injected bug. Concerning \gameTwo, we assume that a performance bug manifests when the agent enters the four \emph{gates} indicated by the white arrows in \figref{fig:games}-(b). 

As detailed in \secref{sub:analysis1}, we assess the extent to which \approach is able to identify the bugs we injected while playing the games. To have a baseline, we compare its results with those of a RL-based agent only trained to play each of the two games (from hereon, \baseline), and with a \random. Since \approach will be trained with the goal of identifying the bugs (details follow), we expect it to adapt its behavior to not only successfully play the game, but to also exercise more often the ``buggy'' areas of the games.

\subsubsection{Learning to Play: RL Models and Game Reward Functions}
We trained the \baseline agent (\ie the one only trained to learn how to play) for \gameOne using the \emph{cross-entropy method} \cite{crossentropy} as RL model. We choose this method because, despite its simplicity, it has been shown to be effective in applications of RL to small environments such as \gameOne \cite{lapan:rlbook}. 

The core of the cross-entropy method is a feedforward neural network (FNN) that takes as input the state of the game and provides as output the action to perform. The state of the game for \gameOne is a vector of dimension 4 containing information about the $x$ coordinate of the pole's center of mass, the pole's speed, its angle with respect to the platform, and its angular speed. There are two possible actions: go right, go left. Once initialized with random weights, the agent (\ie the FNN) starts playing while retaining the experience acquired in each episode: The experience is represented by the state, the action, and the reward obtained during each step of the episode. 

The goal is to keep the pole in balance as long as possible or until the maximum length of an episode (that we set to 1,000 steps) is reached. The \textit{game reward function} is defined so that the agent receives a +1 reward for each step it manages to keep the pole balanced. The total score achieved is also saved at the end of each episode. After $n=16$ consecutive episodes the agent stops playing, selects the $m=11$ (70\%) episodes having the highest score, and uses the experience in those episodes to update the weights of the FNN ($n$ and $m$ have been set according to \cite{lapan:rlbook}). 

Instead, we trained the \baseline agent for \gameTwo using a Deep Q Network (DQN) \cite{mnih2013playing}. In our context, a DQN is a Convolutional Neural Network (CNN) that takes as input a set of contiguous screenshots of the game (in our case 4, as done in previous works \cite{mnih2013playing, mnih2015human}) representing the state of the game and returns, for each possible action defined in the game (five in this case: go up, go right, go down, go left, do nothing), a value indicating the expected reward for the action given the current state (Q value). The multiple screenshots are needed to provide more information to the model about what is happening in the game (\eg in which direction the agent is moving). The goal of the DQN is the same as the FNN: selecting the best action to perform to maximize the reward given the current state. Differently from the previous model, the DQN is updated not on entire episodes but by randomly batching ``experience instances'' among 10k steps saved during the most recent episodes. An ``experience instance'' is saved after each step $\tau$, and is represented by the quadruple ($s_{\tau - 1}, a_{\tau}, s_{\tau}, r_{\tau}$), where $s_{\tau - 1}$ is the input state, $a_{\tau}$ is the action selected by the agent, $s_{\tau}$ is the resulting state obtained by running $a_{\tau}$ in $s_{\tau - 1}$ and $r_{\tau}$ is the received reward. 

The CNN is initialized with random weights, and the agent starts playing while retaining the experience of each step. When enough experience instances have been collected (10k in our implementation \cite{lapan:rlbook}), the CNN starts updating at each step selecting a random batch of experience instances. The reward function for \gameTwo provides a +1 reward every time the agent eats one of the dots and a 0 reward otherwise.

\subsubsection{Instantiating \approach: Performance Reward Functions}
To train \approach to play while looking for the injected bugs, we use a simple \textit{performance reward function}: In both the games, we give a reward of +50 every time the agent, during an episode, spots one of the injected artificial bugs. As previously mentioned, the bugs reveal themselves only the first time the agent visits each buggy position; this means that the performance-based reward is given at most twice for \gameOne and four times for \gameTwo.

\subsubsection{Data Collection and Analysis} \label{sub:analysis1}
We compare \approach against the two previously mentioned baselines: \baseline and the \random. Both \approach and \baseline have been trained for 3,200 episodes on \gameOne and 1,000 on \gameTwo. The different numbers are due to differences in the games and in the RL model we exploited. In both cases, we used a number of episodes sufficient for \baseline to learn how to play (\ie we observed a convergence in the score achieved by the agent in the episodes). 

Once trained, the agents have been run on both games for additional 1,000 episodes, storing the performance bugs they managed to identify in each episode. Since different trainings could result in models playing the game following different strategies, we repeated this process ten times. This means that we trained 10 different models for both \approach and \baseline and, then, we used each of the 10 models to play additional 1,000 episodes collecting the spotted performance bugs. Similarly, we executed \random 10 times for 1,000 episodes each. In this case, no training was needed.

We report descriptive statistics (mean, median, and standard deviation) of the number of performance bugs identified in the 1,000 played episodes by the three approaches. A high number of episodes in which an approach can spot the injected bugs indicate its ability to look for performance bugs while playing the game.

\begin{table*}[t]
	\centering
	\caption{Number of episodes (out of 1,000) in which \emph{RELINE}, \baseline, and the \random identify the injected bugs.}
	\label{tab:results1}
	\begin{tabular}{lrrrrrrrrrrrrr}
		\toprule
		\multirow{2}{*}{\textbf{Game}} & \textbf{\#Injected} & \textbf{\#Bugs} & \multicolumn{3}{c}{\textbf{\emph{RELINE}}} && \multicolumn{3}{c}{\textbf{\emph{rl-baseline}}} && \multicolumn{3}{c}{\textbf{\emph{random agent}}}\\ \cline{4-6} \cline{8-10} \cline{12-14}
		& \textbf{Bugs} & \textbf{found} & \textbf{mean} & \textbf{median} & \textbf{stdev} && \textbf{mean} & \textbf{median} & \textbf{stdev} && \textbf{mean} & \textbf{median} & \textbf{stdev} \\ 
		\midrule
		\multirow{2}{*}{CartPole} & \multirow{2}{*}{2} & 1 & 965 & 984 & 47 && 715 & 706 & 107 && 12 & 11 & 4\\
		& & 2 & 102 & 47 & 177 && 5 & 1 & 7 && 0 & 0 & 0\\
		\midrule
		\multirow{4}{*}{MsPacman} & \multirow{4}{*}{4} 
		& 1 & 971 & 989 & 59 && 700 & 680 & 228 && 24 & 23 & 5\\
		& & 2 & 966 & 985 & 63 && 356 & 343 & 169 && 17 & 16 & 3\\
		& & 3 & 914 & 941 & 87 && 114 & 80 & 90 && 1 & 1 & 1\\  
		& & 4 & 879 & 907 & 106 && 25 & 23 & 17 && 1 & 1 & 1\\    
		\bottomrule
	\end{tabular}
\end{table*}

\subsection{Preliminary Study Results}
\tabref{tab:results1} shows for each of the two games (\gameOne and \gameTwo) the number $k$ of artificial bugs we injected and, for each of the three techniques (\ie~\approach, \baseline, and the \random), descriptive statistics of the number of episodes (out of 1,000) they managed to identify at least $n$ of the injected bugs, with $n$ going from 1 to $k$ at steps of 1.

For both games, it is easy to see that the \random is rarely able to identify the bugs. Indeed, this agent plays without any strategy as it is able to identify bugs only by chance in a few episodes out of the 1,000 it plays. This is also due to the fact that the \random quickly looses the played episodes due to its inability to play the game. This confirms that these approaches are not suitable for testing video games. 

Concerning \gameOne, both \approach and \baseline are able to spot at least one of the two bugs in several of the 1,000 episodes. The median is 984 for \approach and 706 for \baseline. The success of \baseline is soon explained by the characteristics of \gameOne: Considering where we injected the bugs --- see \figref{fig:games}-(a) --- by playing the game it is likely to discover at least one bug (\eg if the player tends to move towards left, the bug on the left will be found). What it is instead unlikely to happen by chance is to find both bugs within the same episode. We found that it is quite challenging, even for a human player, to move the cart first towards one side (\eg left) and, then, towards the other side (right) without losing due to the pole moving more than 12 degrees from vertical. As it can be seen in \tabref{tab:results1}, \approach succeeds in this, on average, for 102 episodes out of 1,000 (median 47), as compared to the 5 (median 1) of \baseline. This indicates that \approach is pushed by the reward function to explore the game looking for the injected bugs, even if this makes playing the game more challenging. Similar results have been achieved on \gameTwo. 

In this case, the DQN is effective in allowing \approach to play while exercising the points in the game in which we injected the bugs. Indeed, on average, \approach was able to spot all four injected bugs in 879 out of the 1,000 played episodes (median=907), while \baseline could achieve such a result only in 25 episodes.

\vspace{0.2cm}
\begin{resultbox}
\textbf{Summary of the Prelimiary Study.} \approach allows obtain agents able not only to effectively play a game but also to spot performance issues. Compared to \baseline, the main advantage of \approach is that it identifies bugs more frequently while playing. 
\end{resultbox}

%% file: study2.tex

\newcommand{\traingameskart}{2,300\xspace}

\section{Case Study: Load Testing an Open Source Game}
\label{sec:study2}
We run a case study to experiment the capability of \approach in load testing an open-source 3D game. Differently from our preliminary study (\secref{sec:study1}), we do not inject artificial bugs. Instead, we aim at finding parts of the game resulting in FPS drops.

\subsection{Study Design}
For this study, we use a 3D kart racing game named \emph{SuperTuxKart} \cite{kart} --- see \figref{fig:games}-(c). This game has been selected due to the following reasons. First, we wanted a 3D game in which, as compared to a 2D game, FPS drops are more likely because of the more complex rendering procedures. Second, SuperTuxKart is popular open-source project that counts, at the time of writing, over 3k stars on GitHub. Third, it is available an open-source wrapper that simplifies the implementation of agents for SuperTuxKart \cite{PySuperTuxKart}. 

The existence of a wrapper like the one we used is crucial since it allows, for example, to advance in the game frame by frame (thus simplifying the generation of the inputs to the RL model), to execute actions (\eg throttle or brake), and to acquire game internals (\eg kart centering, distance to the finish line). 
Also, using this wrapper, it is possible to compute the time needed by the game to render each frame and, consequently, calculate the FPS. Finally, the wrapper allows to have simplified graphics (\eg removing particle effects, like rain, that could make the training more challenging).

\subsubsection{Learning to Play: RL Models and Game Reward Functions} 
The training of the \baseline agent has been performed using the DQN model previously applied in \gameTwo. 

We use the previously mentioned \emph{PySuperTuxKart} \cite{PySuperTuxKart} to make the agent interact with the game. For the sake of speeding up the training, the screenshots extracted from the game have been resized to 200x150 pixels and converted in grayscale before they are provided as input to the model. Moreover, as previously done for \gameTwo, multiple (four) screenshots are fed to the model at each step. Thus, the representation of the state of the game provided to the model is a 4$\times$200$\times$150 tensor. The details of the model and its implementation are available in our replication package \cite{replication}.

A critical part of the learning process is the definition of the \textit{game reward function}. Being SuperTuxKart a racing game, an option could have been to penalize the agent for each additional step required to finish the game. Consequently, to maximize the final score, the agent would have been pushed to reduce the number of steps and, therefore, to drive as fast as possible towards the finish line. However, considering the non-trivial size of the game space, such a reward function would have required a long training time. Thus, we took advantage of the information that can be extracted from the game to help the agent in the learning process. 

SuperTuxKart provides two coordinates indicating where the agent is in the game: \texttt{path\_done} and \texttt{centering}. 

The former indicates the path traversed by the agent from the starting line of the track, while the latter represents the distance of the agent from the center of the track. In particular, \texttt{centering} equals 0 if the agent is at the center of the track, and it moves away from zero as the agent moves to either side: going towards right results in  positive values of the \texttt{centering} value, going left in negative values. We indicate these coordinates with $x$ (\texttt{centering}) and $y$ (\texttt{path\_done}), and we define $\delta_y$ as the path traversed by the agent in a specific step: Given $y_i$ the value for \texttt{path\_done} at step $i$, we compute $\delta_y$ as $y_i - y_{i-1}$. Basically, $\delta_y$ measures how fast the agent is advancing towards the finish line. 

Given $x$ and $\delta_y$ for a given step $i$, we compute the reward function as follows:
$$
\mathit{rg}_i = 
\begin{cases}
 -1 & \text{if }|x| > \theta \\    
  \max(\min(\delta_y, M), 0) & \text{otherwise}
\end{cases}
$$

First, if the agent goes too far from the center of the track ($|x| > \theta$), we penalize it with a negative reward. Otherwise, if the agent is close to the center ($|x| \leq \theta$), we can have two scenarios: (i) if it is not moving towards the finish line ($\delta_y \leq 0$), we do not give any reward (the minimum reward is 0); (ii) if it is moving in the right direction ($\delta_y > 0$), we give a reward proportional to the speed at which it is advancing ($\delta_y$), up to a maximum of M. 

In our experimental setup, we set $\theta = 20$ because it roughly represents the double of $|x|$ when the agent approaches the sides of the road in the level, and $M = 10$ as it is the same maximum reward also given by the \textit{performance reward function}, as we explain below. Finally, we reward the agent when it crosses the finish line with an additional $+1,000$ bonus.

\subsubsection{Instantiating \approach: Performance Reward Function} \label{sub:approachkart}
To define the \textit{performance reward function} of \approach for SuperTuxKart, the first step to perform is to define a way to reliably capture the FPS of the game during the training. In this way, we can reward the agent when it manages to identify low-FPS points. As previously said, we use PySuperTuxKart to interact with the game and such a framework keeps the game frozen while the other instructions of \approach (\eg the identification of the action to execute) are run. Since the framework runs the game in the same process in which we run \approach and since we do not use threads, we can safely use a simple method for computing the time needed to render the four frames: We get the system time before ($T_{\mathit{before}}$) and after ($T_{\mathit{after}}$) we trigger the rendering of the frames and we compute the time needed at step $i$ as $\mathit{rT}_i = T_{\mathit{after}} - T_{\mathit{before}}$. Such a value is negatively correlated with the FPS (higher rendering time means lower FPS).

The \textit{performance reward function} we use is the following:
$$
\mathit{rp}_i = 
\begin{cases}
  10 & \text{if } |x| \leq \theta \land \mathit{rT}_i > t \\
  0 & \text{otherwise}
\end{cases}
$$
We give a performance-based reward of 10 when the agent takes more than $t$ milliseconds to render the frames at a given point (causing an FPS drop). We explain the tuning of $t$ later. We do not give such a reward when $|x| > \theta$ (the kart is far from the center) since we want the agent to spot issues in positions that are likely to be explored by real players (\ie reasonably close to the track).

Finally, in \approach we do not give a fixed $+1,000$ bonus reward when the agent crosses the finish line but we assign a bonus computed as $10 \times \sum_{i = 1}^{\mathit{steps}} rp_{i}$, \ie proportional to the total performance-based reward accumulated by the agent in the episode. 
This is done to push the agent to visit more low-FPS points during an episode.

\subsubsection{Data Collection and Analysis} 
As done in our preliminary study, we compare \approach with \baseline (\ie the agent only trained to play the game) and with a \random. 

\textbf{Training \baseline and \emph{RELINE}.} While we used different reward functions for the two RL agents, we applied the same training process for both of them. We trained each model for \traingameskart episodes, with one episode having a maximum duration of 90 seconds or ending when the agent crosses the finish line of the racing track (the agent is required to perform a single lap). We set the 90 seconds limit since we observed that, by manually playing the game, $\sim$70 seconds are sufficient to complete a lap. The \traingameskart episodes threshold has been defined by computing the average reward obtained by the two agents every 100 episodes and by observing when a plateau was reached by both agents. We found \traingameskart episodes to be a good compromise for both agents (graphs plotting the reward function are available in the replication package \cite{replication}). 

The trained \baseline agent has been used to define the threshold $t$ needed for the \approach's reward function (\ie for identifying when the agent found a low-FPS point and should be rewarded).

In particular, once trained, we run \baseline for 300 episodes, storing the time needed by the game to render the subsequent four frames after every action recommended by the model.\footnote{Since we wanted to measure the frames rendering time in a standard scenario in which the agent was driving the kart, we stopped an episode if the agent got stuck against some obstacle.} This resulted in a total of 48,825 data points $s_{FPS}$, representing the standard FPS of the game in a scenario in which the player is only focused on completing the race as fast as possible. 

Starting from the 48,825 $s_{FPS}$ data points collected in the 300 episodes played by the trained \baseline agent, we apply the five-$\sigma$ rule \cite{grafarend2006linear} to compute a threshold able to identify outliers. The five-$\sigma$ rule states that in a normal distribution (such as $s_{FPS}$) 99.99\% of observed data points lie within five standard deviations from the mean. Thus, anything above this value can be considered as an outlier in terms of milliseconds needed to render the frames. For this reason, we compute $t_b = mean(s_{FPS}) + 5 \times sd(s_{FPS})$ as a candidate base threshold to identify low-FPS points. However, $t_b$ cannot be directly used as the $t$ value of our reward function. Indeed, we observed that the time needed for rendering frames during the \approach's training is slightly higher as compared to the time needed when the trained \baseline agent is used to play the game. This is due to the fact that the load on the server (and in particular on the GPU) is higher during training. To overcome this issue, we perform the following steps. 

At the beginning of the training, we run 100 \emph{warmup episodes} in which we collect the time needed to render the four frames after each action performed by the agent. Then, we compute the first ($Q^{\mathit{tr}}_{1}$) and the third ($Q^{\mathit{tr}}_{3}$) quartile of the obtained distribution and compare them to the $Q_1$ and $Q_3$ of the distribution obtained in the 300 episodes used to define $t_b$ (\ie those played by the trained \baseline agent). 
During the \textit{warmup episodes}, the agent selects the action to perform almost randomly (it still has to learn): Therefore, it would not be able to explore a substantial area of the game (\ie of the racing track), thus not providing a distribution of timings comparable with the ones obtained when the trained \baseline agent that played the 300 episodes. For this reason, during the 100 \emph{warmup episodes} of the training, the action to perform is not chosen by the agent currently under training, but by the trained \baseline agent (\ie the same used in the 300 episodes). This does not impact in any way the load on the server that remains the one we have during the training of \approach since the only change we have is to ask for the action to perform to the \baseline agent rather than to the one under training. However, the whole training procedure (\eg capturing the frames and updating the network) stays the same.  

We compute the additional ``cost'' brought by the training in rendering the frames during the game using the formula $\delta = max(Q^{\mathit{tr}}_{1} - Q_{1}, Q^{\mathit{tr}}_{3} - Q_{3})$. 
We use the first and third quartiles since they represent the boundaries of the central part of the distribution, \ie they should be quite representative of the values in it. 
We took as $\delta$ the maximum of the two differences to be more conservative in assigning rewards when the agent identifies low-FPS points. 
The final value $t$ we use in our reward function when training \approach to load test SuperTuxKart is defined as: $t = t_b + \delta = 18.36$.\footnote{We identify as low-FPS points the ones in which the FPS is lower than 218. Such a number is still very high, more than enough for any human player, in practice. Note that we run the game using high-performance hardware and, most importantly, with the lowest graphic settings. The equivalent in normal conditions would be much lower.} 

Thus, if \approach is able, during the training, to identify a point in the game requiring more than $t$ milliseconds to render four frames, then it receives a reward as explained in \secref{sub:approachkart}.

The training of \baseline took $\sim$3 hours, while \approach requires substantially more time due to the fact that, after each step performed by the agent, we collect and store information about the time needed to render the frames (this is done million of times). This pushed the training for \approach up to $\sim$30 hours.

\textbf{Reliability of Time Measurements.} It is important to clarify that the FPS of the game can be impacted by the hardware specifications and the current load of the machine running it. In other words, running the same game on two different machines or on the same machine in two different moments can result in variations of the FPS. For this reason, all the experiments have been performed on the same server, equipped with 2 x 64 Core AMD 2.25GHz CPUs, 512GB DDR4 3200MHz RAM, and an nVidia Tesla V100S 32GB GPU. Also, the process running the training of the agents or the collection of the 48,825 $s_{FPS}$ with the trained \baseline agent was the only process running on the machine besides those handled by the operating system (Ubuntu 20.04). On top of that, the process was always run using the \texttt{chrt --rr 1} option, that in Linux maximizes the priority of the process, reducing the likelihood of interruptions. 

Despite these precautions, it is still possible that variations are observed in the FPS not due to issues in the game, but to external factors (\eg changes in the load of the machine). To verify the reliability of the collected FPS data, we run a constant agent performing always the same actions in the game for 300 episodes. The set of actions has been extracted from one of the episodes played by the \baseline agent, that was able to successfully conclude the race. Then, we plotted the time needed by the game to render the four frames following each action made by the agent. Since we are playing 300 times exactly the same episode, we expect to observe the same trend in terms of FPS for each game. If this is the case, it means that the way we are measuring the FPS is reliable enough to reward the agent when low-FPS points are identified. 

\begin{figure}[t]
	\centering
	\includegraphics[width=\linewidth]{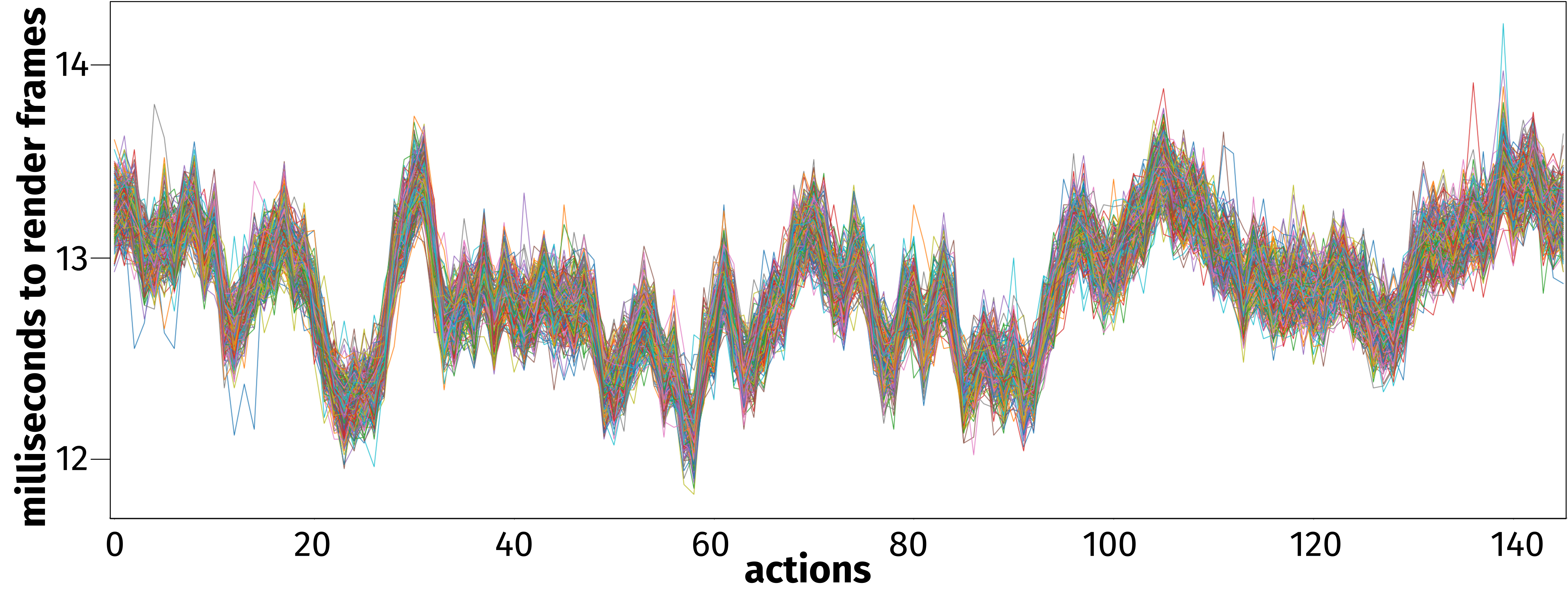}
	\caption{Rendering times for 300 episodes (same actions).}
	\label{fig:stability}
\end{figure}

\figref{fig:stability} shows the achieved results: The $y$-axis represents the milliseconds needed to render four frames in response to an agent's action ($x$-axis) performed in a specific part of the game. While, as expected, small variations are possible, the overall trend is quite stable: Points of the game requiring longer time to render frames are consistently showing across the 300 episodes, resulting in a clear trend. We also computed the Spearman's correlation \cite{spearman04} pairwise across the 300 distributions, adjusting the obtained $p$-values using the Holm's correction \cite{Holm1979a}. 

We found all correlations to be statistically significant (adjusted $p$-values $<$ 0.05) with a minimum $\rho$=0.77 (strong correlation) and a median $\rho$=0.91 (very strong correlation). This confirms the common FPS trends across the 300 episodes.

\textbf{Running the Three Techniques to Spot Low-FPS Areas.} After the \traingameskart training episodes, we assume that both the RL-based agents learned how to play the game, and that \approach also learned how to spot low-FPS points. Then, as also done in our preliminary study, we train both agents for additional 1,000 episodes, storing the time needed to render the frames in every single point they explored during each episode (where a point is represented by its coordinates, \ie \texttt{centering}=$x$ and \texttt{path\_done}=$y$). We do the same also with the \random. 

\textbf{Data Analysis.} The output of each of the three agents is a list of points with the milliseconds each of them required to render the subsequent frames. Since each agent played 1,000 episodes, it is possible that the same point is covered several times by an agent, with slightly different FPS observed (as previously explained, small variations in FPS are possible and expected across different episodes). We classify as low-FPS points those that required more than $t$ milliseconds to render the four subsequent frames more than 50\% of times they have been covered by an agent. 

This means that, if across the 1,000 episodes a point $p$ is exercised 100 times by an agent, at least 51 times the threshold $t$ must be exceeded to consider $p$ as a low-FPS point. In practice, a developer using \approach for identifying low-FPS points could use a higher threshold to increase the reliability of the findings. However, for the sake of this empirical study, we decided to be conservative.

\begin{figure}[tb]
\centering
\includegraphics[width=\linewidth]{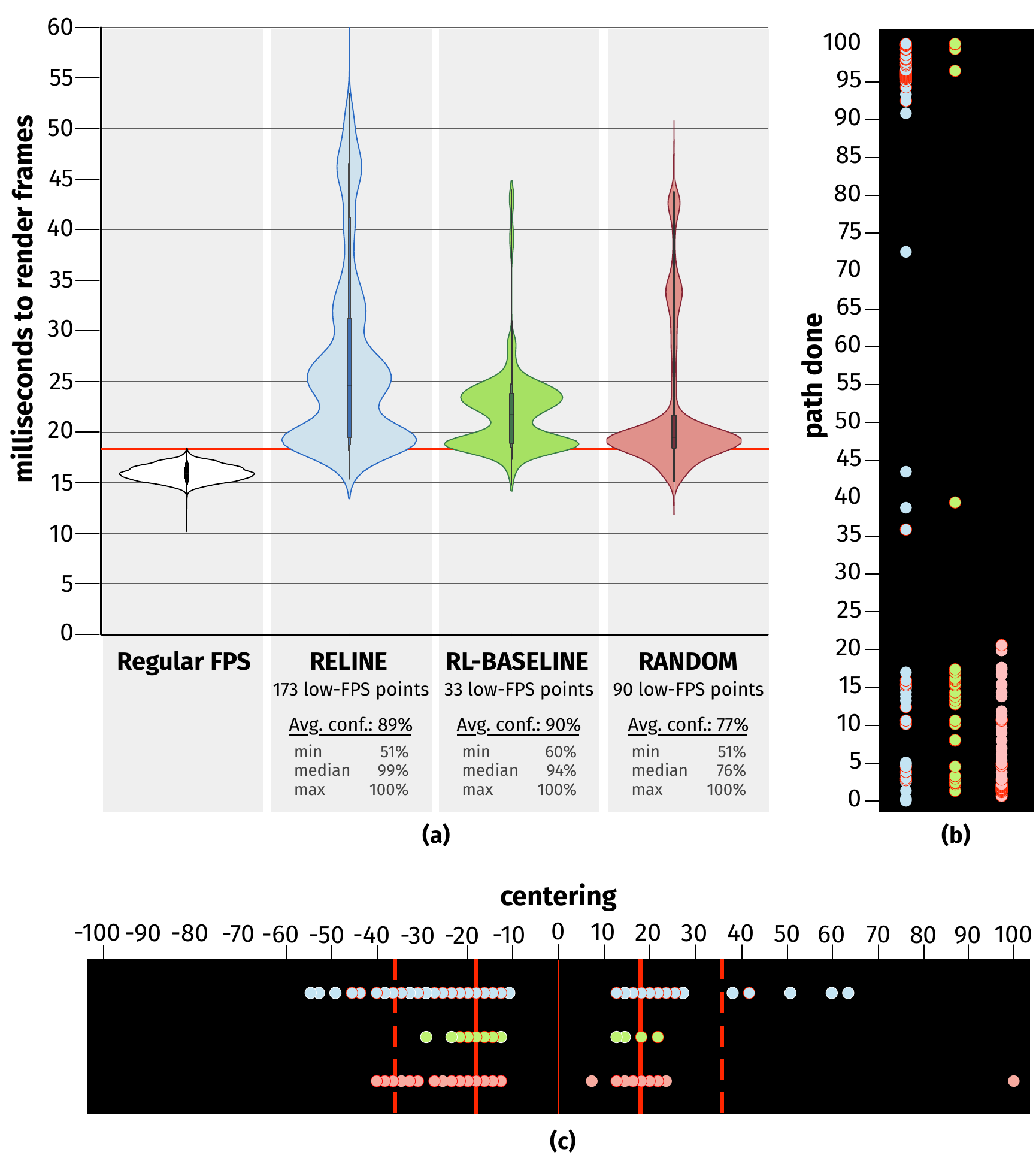}
\caption{Results of the study: (a) reports the distributions of timings for the low-FPS points with summary statistics, while (b) and (c) depict the \textit{path done} and \textit{centering} coordinates at which the such points were observed, respectively.}
\label{fig:results2}
\end{figure}

Then, we compare the characteristics of the low-FPS points identified by the three approaches. Specifically, we analyze: (i) how many different low-FPS points each approach identified; (ii) the number of times each low-FPS point has been exercised by each agent in the 1,000 episodes; (iii) the \emph{confidence} of the identified points (\ie the percentage of times an exercised point resulted in low FPS). Given the low-FPS points identified by each agent, we draw violin plots showing the distribution of timings needed to render the frames when the agent exercised them (the higher the timings, the lower the FPS). We compare these distributions using Mann-Whitney test \cite{Conover:1998} with $p$-values adjustment using the Holm's correction \cite{Holm1979a}. We also estimate the magnitude of the differences by using the Cliff's Delta ($d$), a non-parametric effect size measure \cite{Cliff:2005} for ordinal data. We follow well-established guidelines to interpret the effect size: negligible for $|d| < 0.10$, small for $0.10 \le |d| < 0.33$, medium for $0.33 \le |d| < 0.474$, and large for $|d| \ge 0.474$ \cite{Cliff:2005}.

\subsection{Study Results}
\figref{fig:results2} summarizes the main findings of our case study. \figref{fig:results2}-(a) shows the distribution of time needed to render the game frames (\ie our proxy for FPS) for four groups of points. The first violin plot on the left (\ie Regular FPS) shows the timing for points that have never resulted in a drop of FPS in any of the 3,000 episodes played by the three agents (1,000 each). These serve as baseline to better interpret the low-FPS points exercised by the agents. The other three violin plots show the distributions of timing for the low-FPS points identified by \approach (blue), \baseline (green), and the \random (red). 

Below each violin plot we report the number of low-FPS points identified by each agent and descriptive statistics (average, median, min, max) of the confidence for the low-FPS points. A 100\% confidence means that all times that a low-FPS point has been exercised in the 1,000 episodes played by the agent it required more than $t=18.36$ milliseconds to render the subsequent frames. The $t$ threshold is represented by the red horizontal line. On average, \approach exercised each low-FPS point 89 times in the 1,000 episodes, against the 210 of \baseline and the 829 of the \random (the same point can be exercised multiple times in an episode).

\approach identified 173 low-FPS points, as compared to the 33 of \baseline and the 90 of the \random. The confidence is similar for \approach (median=99\%) and \baseline (median=94\%), while it is lower for the \random (median=76\%). Thus, the low-FPS points identified by the two RL-based agents are, overall, quite reliable. Concerning the number of low-FPS points identified, \approach identifies more points as compared to \baseline (173 \emph{vs} 33). This is expected since it has the explicit goal of load testing the game, However, what could be surprising at first sight is the high number of low-FPS points identified by the \random (90). \figref{fig:results2}-(b) and \figref{fig:results2}-(c) help in interpreting this finding.

\figref{fig:results2}-(b) plots the \texttt{path\_done} ($y$ coordinate) for each low-FPS point identified by each agent, using the same color schema of the violin plots (\eg blue corresponds to \approach). 

If multiple points fall in the same coordinate (\ie same \texttt{path\_done} but different \texttt{centering}), they are shown with a red border. The scale of the \texttt{path\_done} has been normalized between 0 and 100, where 0 corresponds to the starting line of the track and 100 to its finish line. Similarly, \figref{fig:results2}-(c) plots the \texttt{centering} ($x$ coordinate) for the low-FPS points. The line at 0 represents the center of the track, while the continuous lines in position $\sim$-18 and $\sim$18 depict the limits of the track. Finally, the dashed lines represent the area of the game we asked \approach to explore: based on our reward function, we penalize the agent for going outside the [-20, +20] range that, normalized, corresponds to $\sim$[-36, +36]. Also \baseline is penalized outside of this area.

As expected, the \random is not able to advance in the game: The low-FPS points it identifies are all placed near the starting line --- red dots in \figref{fig:results2}-(b). This indicates that a random agent can be used to exercise a specific part of a game, but it is not able to explore the game as a player would do. This is also confirmed by the red dots in \figref{fig:results2}-(c), with the \random exploring areas of the game far from the track and that a human player is unlikely to explore. Also, it is worth noting that in SuperTuxKart each episode lasts (based on our setting) 90 seconds if the agent does not cross the finish line. However, as shown in our preliminary study, in other games such as \gameTwo a \random could quickly lose an episode without having the chance to explore the game at all. 

The low-FPS points identified by \approach (blue dots) and by \baseline (green) are instead closer to the track and, for what concerns \approach, they are within or very close the area of the game we ask it to explore --- see dashed lines in \figref{fig:results2}-(c). Thus, by customizing the reward function, it is possible to define the area of the game relevant for the load testing.

Looking at \figref{fig:results2}-(b), we can see that \approach is also able to identify low-FPS in different areas of the game with, however, a concentration close to the beginning and the end of the game. It is difficult to explain the reason for such a result, but we hypothesize two possible explanations. 

First, it is possible that the ``central'' part of the game simply features less low-FPS areas. This would also be confirmed by the fact that \baseline only found one low-FPS point in that part of the game. Also, the training and the reward function could have driven \approach to explore more the starting and the ending of the game. The starting part is certainly the most explored since, at the beginning of the training, the agent is basically a random agent. Thus, it mostly collects experience about low-FPS points found in the beginning of the game since, similarly to the \random, it is not able to advance in the game. It is important to remember that the data in \figref{fig:results2} only refers to the 1,000 games played by \approach after the 2,300 training games, so we are not including the random exploration done at the beginning of the training in \figref{fig:results2}. However, once the agent learns several low-FPS points in the starting of the game, it can exercise them again and again to get a higher reward. 

Concerning the end of the game, we set a maximum duration of 90 seconds for each game, but we know that a well-trained agent can complete the lap in $\sim$70 seconds. It is possible that the agent used the remaining time to better explore the last part of the game before crossing the finish line, thus finding a higher number of low-FPS points in that area. Additional trainings, possibly with a different \emph{reward function}, are needed to better explain our finding.

Concerning the violin plots in \figref{fig:results2}-(a), we can see that \approach and \baseline exhibit a similar distribution, with \approach being able to identify some stronger low-FPS points (\ie longer time to render frames). All distributions have, as expected, the median above the $t$ threshold, with \approach's one being higher (24.54 \emph{vs} 21.69 for \baseline and 19.39 for \random). The highest value of the distributions is 65.92 (60.7 FPS) for \approach, against 44.81 (89.3 FPS) for \baseline and 50.73 (78.8 FPS) for \random. Remember that all these values represent milliseconds to load frames after an action performed by the agents.

\begin{table}[ht]
	\centering
	\caption{Results of Mann-Whitney test (adjusted $p$-value) and Cliff's Delta ($d$) when comparing the distributions of rendering times --- boldface indicates higher times.}
        \label{tab:stats}
	\begin{tabular}{lrr}
		\toprule
		\textbf{Test} & \textbf{\emph{p}-value} & \textbf{OR} \\ 
		\midrule 
		\textbf{\emph{RELINE}} $vs$ \baseline & $<$0.001 & 0.34 (Medium)\\
		\textbf{\emph{RELINE}} $vs$ \random & $<$0.001 & 0.36 (Medium)\\
		\textbf{\baseline} $vs$ \random & $<$0.001 & 0.16 (Small)\\		
		\bottomrule
	\end{tabular}
\end{table}

\tabref{tab:stats} shows the results of the statistical comparisons among the three distributions. In each test, the approach reported in boldface is the one identifying stronger low-FPS points (\ie more extreme points requiring longer rendering time for their frames). The adjusted $p$-values report a significant difference ($p$-value $<$ 0.001) in favor of \approach against both \baseline and the \random (in both cases, with a medium effect size). Thus, the low-FPS points identified by \approach tend to require longer times to render frames. \figref{fig:games}-(c) shows an example of low-FPS point identified by \approach: Crashing against the sheep results in a drop of FPS.

Finally, it is worth commenting about the overlap of low-FPS points identified by the three agents. Indeed, \approach and \baseline found 14 low-FPS points in common (\ie same $x$ and $y$ coordinates), while the overlap is of 11 points for \approach and \random, and 10 for \baseline and \random. The most interesting finding of this analysis is that \baseline was able to identify only 19 points missed by \approach, while the latter found 159 points missed by \baseline. This supports the role played by the \emph{reward function} in pushing \approach to look for low-FPS points.

\vspace{0.2cm}
\begin{resultbox}
 \textbf{Summary of the Case Study.} \approach is the best approach for finding low-FPS points in SuperTuxKart. A \random is not able to spot issues that require playing skills, and \baseline only finds a small portion of the low-FPS points.
\end{resultbox}

%% file: threats.tex

\section{Threats to Validity}
\label{sec:threats}

\textbf{Threats to Construct Validity}. The main threats to the construct validity of our study are related to the process we adopted in our case study (\secref{sec:study2}) to identify low-FPS points. Based on our experiments, and in particular on the findings reported in \figref{fig:stability}, our methodology should be reliable enough to identify variations in FPS. Still, some level of noise can be expected, and for this reason all our analyses have been run at least 300 times, while 1,000 episodes were played by each of the experimented approaches.

Concerning our preliminary study (\secref{sec:study1}), it is clear that the bugs we injected are not representative of real performance bugs in the subject games. However, they are inspired from a performance mutation operator defined in the literature \cite{performance-mutation}. Our preliminary study only serves as a proof-of-concept to verify whether, by modifying the reward function, a RL-based agent would adapt its behavior to look for bugs while playing the game.

\textbf{Threats to Internal Validity.} In our case study, to ease the training we did not use the ``real'' game, but its wrapped version, \ie PySuperTuxKart \cite{PySuperTuxKart}. While the core game is the same, the version we adopted does not contain the latest updates and it includes additional Python code that may affect the rendering time. We assume that such a time is constant for all the frames since it simply triggers the frame rendering operation in the core game. Besides, we forced the game to run with lowest graphics settings to speed up rendering: For example, we excluded dynamic lighting, anti-aliasing, and shadows. Therefore, the low-FPS points found in PySuperTuxKart may be irrelevant in the original game or with other graphic settings. Also, we applied the five-$\sigma$ rule to define a threshold for defining what a low-FPS point is. The threshold we set might be not indicative of relevant performance issues. 

Still, the goal of our study was to show that once set specific requirements (\eg the threshold $t$, the area to explore, \etc), the agent is able to adapt trying to maximize its reward. Thus, we do not expect changes in the threshold to invalidate our findings.

\textbf{Threats to conclusion validity.} In our data analysis we used appropriate statistical procedures, also adopting \emph{p}-value adjustment when multiple tests were used within the same analysis.

\textbf{Threats to External Validity} Besides the proof-of-concept study we presented in \secref{sec:study1}, our empirical evaluation of \approach includes a single game. This does not allow us to generalize our findings. The reasons for such a choice lie in the high effort we experienced as researchers in (i) building the pipeline to interact with the game, (ii) finding and experimenting with a reliable way to capture the FPS, (iii) defining a meaningful reward function that allowed the agent to successfully play the game in the first place and, then, to also spot low-FPS points. These steps were a long trial-and-error process with the most time consuming part being the trainings needed to test the different reward functions we experimented before converging towards the ones presented in this paper. Indeed, testing a new version of a reward function required at least one week of work with the hardware at our disposal (including implementation, training, and data analysis). 

This was also due to the impossibility of using multiple machines or to run multiple processes in parallel on the same server. Indeed, as explained, using the exact same environment to run all our experiments was a study requirement. It is worth noting that, because of similar issues, other state-of-the-art approaches targeting different game properties were experimented with only one game as well (see \eg \cite{Zook2014AutomaticPF, pfau2020dungeons, Bergdahl2020, wu2020regression}). We believe that instantiating \approach on a new game would be much easier by collaborating with the game developers. While this would only slightly simplify the definition of a meaningful reward function, the original developers of the game could easily provide through APIs all information needed by \approach (including, \eg the FPS), cutting away weeks of work. 

%% file: related.tex

\section{Related Work}
\label{sec:related}
Three recent studies \cite{politowski2021survey, truelove2021we, li2021data} suggest that finding performance issues in video games is a relevant problem, according to both game developers \cite{politowski2021survey, truelove2021we} and players \cite{li2021data}. Nevertheless, to the best of our knowledge, no previous work introduced automated approaches for load testing video games. Therefore, in this section, we discuss some important works on the quality assurance of video games in general. We first introduce the approaches defined in the literature for training agents able to automatically play and win a game. Then, we show how such approaches are used for play-testing for (i) finding functional issues and (ii) assessing game/level design (\eg finding unbalanced levels or mechanics).


\subsection{Training Agents to Play}
Reinforcement Learning (RL) is widely used to train agents able to automatically play video games. Mnih \etal \cite{mnih2013playing, mnih2015human} presented the first approach based on high-dimensional sensory input (\ie raw pixels from the game screen) able to automatically learn how to play a game. The authors used a Convolutional Neural Network (CNN) trained with a variant of Q-learning to train their agent. The proposed approach is able to surpass human expert testers in playing some games from the Atari 2600 benchmark.

Vinyals \etal \cite{Vinyals2017StarCraftIA} introduced SC2LE, a RL environment based on the game \textit{StarCraft II} that simplifies the development of specialized agents for a multi-agent environment.

Hessel \etal \cite{hessel2018rainbow} analyzed six extensions of the DQN algorithm for RL and they reported the combinations that allow to achieve the best results in terms of training time on the Atari 2600 benchmark.

Baker \etal \cite{baker2019emergent} explored the use of RL in a multi-agent environment (\ie the \textit{hide and seek} game). They report that agents create self-supervised autocurricula \cite{leibo2019autocurricula}, \ie curricula naturally emerging from competition and cooperation. 
As a result, the authors found evidence of strategy learning not guided by direct incentives.

Berner \etal \cite{berner2019dota} reported that state-of-the-art RL techniques were successfully used in OpenAI Five to train an agent able to play Dota 2 and to defeat the world champion in 2019 (Team OG). Finally, Mesentier \etal \cite{Mesentier:2017} reported that AI agents could be easily trained to explore the states of a board game (\textit{Ticket to Ride}) performing automated play-testing.

\subsection{Testing of Video Games}
Functional testing of video games aims at finding unexpected behaviors in a game. Defining the test oracle, \ie determining if a specific game behavior is defective, is not trivial. Several categories of test oracles were identified to determine if a bug was found: \textit{crash} (the game stops working) \cite{Pfau:2017, Zheng:ase2019}, \textit{stuck} (the agent can not win the game) \cite{Pfau:2017, Zheng:ase2019}, \textit{game balance} (game too easy or too hard) \cite{Zheng:ase2019}, \textit{logical} (an invalid state is reached) \cite{Zheng:ase2019}, and \textit{user experience bugs} (related to graphic and sound, \eg glitches) \cite{Pfau:2017, Zheng:ase2019}. While heuristics can be used to find possible crash-, stuck-, and game-balance-related bugs \cite{Zheng:ase2019}, logical and user-experience bugs may require the developers to manually define an oracle.

Iftikhar \etal \cite{Iftikhar:models2015} proposed a model-based testing approach for automatically perform black-box testing of platform games. More recent approaches mostly rely on RL.
 
Pfau \etal \cite{Pfau:2017} introduced ICARUS, a framework for autonomous play-testing aimed at finding bugs. ICARUS supports the fully automated detection of \textit{crash} and \textit{stuck} bugs, while it also provides semi-supervised support for \textit{user-experience} bugs.
 
Zheng \etal \cite{Zheng:ase2019} used Deep Reinforcement Learning (DLR) in their approach, Wuji. Wuji balances the aim of winning the game and exploring the space to find \textit{crash}, \textit{stuck}, \textit{game balance}, and \textit{logical} bugs in three video games (one simple, \textit{Block Maze} and two commercial, \textit{L10} and \textit{NSH}).

Bergdahl \etal \cite{Bergdahl2020} defined a DLR-based method which provides support for continuous actions (\eg mouse or game-pads) and they experimented it with a first-person shooter game.

Wu \etal \cite{wu2020regression} used RL to automatically perform regression testing, \ie to compare the game behaviors in different versions of a game. They experimented with such an approach on a Massive Multiplayer Online Role-Playing Game (MMORPG).

Ariyurek \etal \cite{Ariyurek:tg2021} experimented RL and Monte Carlo Tree Search (MCTS) to define both synthetic agents, trained in a completely automated manner, and human-like agents, trained on trajectories used by human testers.

Finally, Ahumada and Bergel \cite{ahumada2020reproducing} proposed an approach based on genetic algorithms to reproduce bugs in video games by reconstructing the correct sequence of actions that lead to the desired faulty state of the game.

\subsection{Game- and Level-Design Assessment}
One of the main goals of a video game is to provide a pleasant gameplay to the player. Assessing the game balance and other aspects related to game- and level-design is, therefore, of primary importance. 

For this reason, previous work defined several approaches for automatically finding game- and level-design issues in video games.
Zook \etal \cite{Zook2014AutomaticPF} proposed an approach based on Active Learning (AL) to help designers performing low-level parameter tuning. They experimented such an approach on a \textit{shoot 'em up} game.

Gudmundsson \etal \cite{gudmundsson2018human} introduced an approach based on Deep Learning to learn human-like play-testing from player data. They used a CNN to automatically predict the most natural next action a player would take aiming to estimate difficulty of levels in \textit{Candy Crush Saga} and \textit{Candy Crush Soda Saga}.

Zhao \etal \cite{Zhao2019} report four case studies in which they experiment the use of human-like agent trained with RL to predict player interactions with the game and to highlight possible game-design issues.
On a similar note, Pfau \etal \cite{pfau2020dungeons} used deep player behavioral models to represent a specific player population for \textit{Aion}, a MMORPG. They used such models to estimate the game balance and they showed that they can be used to tune it.

Finally, Stahlke \etal \cite{Stahlke2020ArtificialPI} defined PathOS, a tool aimed at helping developers to simulate players' interaction with a specific game level, to understand the impact of small design changes.


%% file: conclusion.tex

\section{Conclusions and Future Work}
\label{sec:conclusions}
We presented \approach, an approach that uses RL to load test video games. \approach can be instantiated on different games using different RL models and reward functions. 

Our proof-of-concept study performed on two subject systems shows the feasibility of our approach: Given a reward function able to reward the agent when artificial performance bugs are identified, the agent adapts its behavior to play the game while looking for those bugs. 

We performed a case study on a real 3D racing game, SuperTuxKart, showing the ability of \approach to identify areas resulting in FPS drops. As compared to a classic RL agent only trained to play the game, \approach is able to identify a substantially higher number of low-FPS points (173 \emph{vs} 33).

Despite the encouraging results, there are many aspects that deserve a deeper investigation and from which our future research agenda stems. First, we plan additional tests on SuperTuxKart to better understand how the agent reacts to changes in the reward function (\eg is it possible to find more low-FPS points in the central part of the game?). Also, with longer training times it should be possible to train an agent able to play more challenging versions of this game featuring additional 3D effects (\eg rainy conditions), possibly allowing to find new low-FPS points. We also plan to instantiate \approach on other game genres (\eg role-playing games), possibly by cooperating with their developers. 

In our replication package \cite{replication}, we release the code implementing the models used in our study and the raw data of our experiments.